\documentclass[11pt]{article}
\usepackage{booktabs}
\usepackage{PRIMEarxiv}
\usepackage{graphicx}
\usepackage{url}
\usepackage{microtype}
\usepackage{tabularx}
\usepackage{amsmath}
\usepackage[colorlinks=true,linkcolor=blue,citecolor=blue,urlcolor=blue]{hyperref}

\newcommand{\pattack}{\ensuremath{p(\mathrm{attack})}}
\newcommand{\fnr}{\mathrm{FNR}}

\title{From Detection to Decision: Calibration and Deferral for Prompt-Injection Detection}
\author{
Qianlong Lan, Anuj Kaul, Vinothini Pandurangan, Indranil Sanyal \\
 eBay Inc. \\
}
\date{}
\begin{document}
\maketitle

\begin{abstract}
Prompt-injection defenses are commonly evaluated as binary detectors, but deployed systems must decide whether to allow, block, or escalate an input. Recent decision models claim to provide probabilities suitable for making these decisions automatically. We examine whether these probabilities lead to better security decisions. We compare Jev, a native probabilistic decision model, with Llama Prompt Guard 2 classifiers (22M and 86M parameters) and a frontier LLM judge. Our evaluation uses 47,611 held-out confirmatory examples covering human-written direct attacks, indirect injections, hard-benign prompts, obfuscation, and distribution shift. Prompt and calibration choices are determined on separate development and calibration splits and fixed before confirmatory evaluation. The LLM judge achieves the highest overall discrimination (AUROC 0.897), followed by Jev (0.871), Prompt Guard 2 86M (0.757), and Prompt Guard 2 22M (0.714). Jev's native probabilities are not consistently better calibrated than temperature-scaled classifier scores: Jev obtains ECE 0.541, Brier 0.440, and NLL 1.432, compared with 0.571, 0.439, and 1.104 for calibrated Prompt Guard 2 86M. On the frozen confirmatory cohort, the empirical deferral curves satisfy $\fnr\leq1\%$ only when more than 91.8\% of inputs are escalated, with substantial false-positive rates. This crossing is a descriptive confirmatory statistic, not a deployment threshold chosen using test labels. These results indicate that training for calibrated outputs alone does not ensure reliable security decisions under distribution shift. Discrimination, calibration, reviewer behavior, and policy utility need to be evaluated jointly.
\end{abstract}

\keywords{Prompt Injection Detection \and Probability Calibration \and Selective Prediction \and Uncertainty Estimation \and Distribution Shift \and Large Language Models}

\section{Introduction}

Prompt injection turns untrusted text into instructions that redirect an LLM application. A detector can assign a label, but a deployed system must make a decision: allow a low-risk input, block a high-risk one, or defer an uncertain case to a more capable reviewer. This decision depends on what the detector's score represents. A probability of 0.9 is useful only if high scores reliably indicate attacks and remain meaningful as attacks, domains, or benign prompts change.

Prior work has introduced a range of prompt-injection benchmarks. TensorTrust contains human-authored adversarial prompts~\cite{toyer2024tensortrust}; BIPIA targets indirect injection through external content~\cite{yi2023bipia}; NotInject measures over-defense on benign prompts containing attack-like terms~\cite{li2024injecguard}; and PIDS-Bench evaluates robustness to obfuscation and distribution shift~\cite{shire2026pids}. Recent work has also examined detector calibration under attack shift~\cite{confidentlywrong2026}. In parallel, TypeSafe describes Jev as a typed decision model trained with Reinforcement Learning for Calibrated Decisions (RLCD) to produce probabilities that software can use for routing and escalation~\cite{typesafe2026jev}. These developments motivate a direct question: does a model trained for calibrated decisions make better security decisions than a specialized classifier with conventional post-hoc calibration?

We study three research questions. 
\textbf{RQ1} compares the detection performance of Jev, Prompt Guard 2, and an LLM judge. 
\textbf{RQ2} examines whether the native probabilities of an off-the-shelf decision model are more reliable on the target workload than raw classifier scores or classifier scores adjusted with conventional post-hoc calibration, particularly under distribution shift. \textbf{RQ3} examines how uncertainty at the gate translates into ALLOW--BLOCK--DEFER decisions when paired with a frozen imperfect reviewer. We measure this behavior by the empirical confirmatory escalation rate required to achieve system $\fnr\leq1\%$.

We make three contributions. We provide a leakage-controlled comparison of a native decision model, two specialized classifiers, and a verbalized-probability LLM judge on 47,611 confirmatory examples. We jointly evaluate discrimination, probability calibration, confident false negatives, latency, and operational security-deferral behavior. Our results show that native decision probabilities do not consistently outperform target-calibrated classifier probabilities across calibration metrics, and that the evaluated gate--reviewer systems reach a strict 1\% FNR target only at very high confirmatory escalation rates.

\begin{figure}[t]
  \centering
  \includegraphics[width=\linewidth]{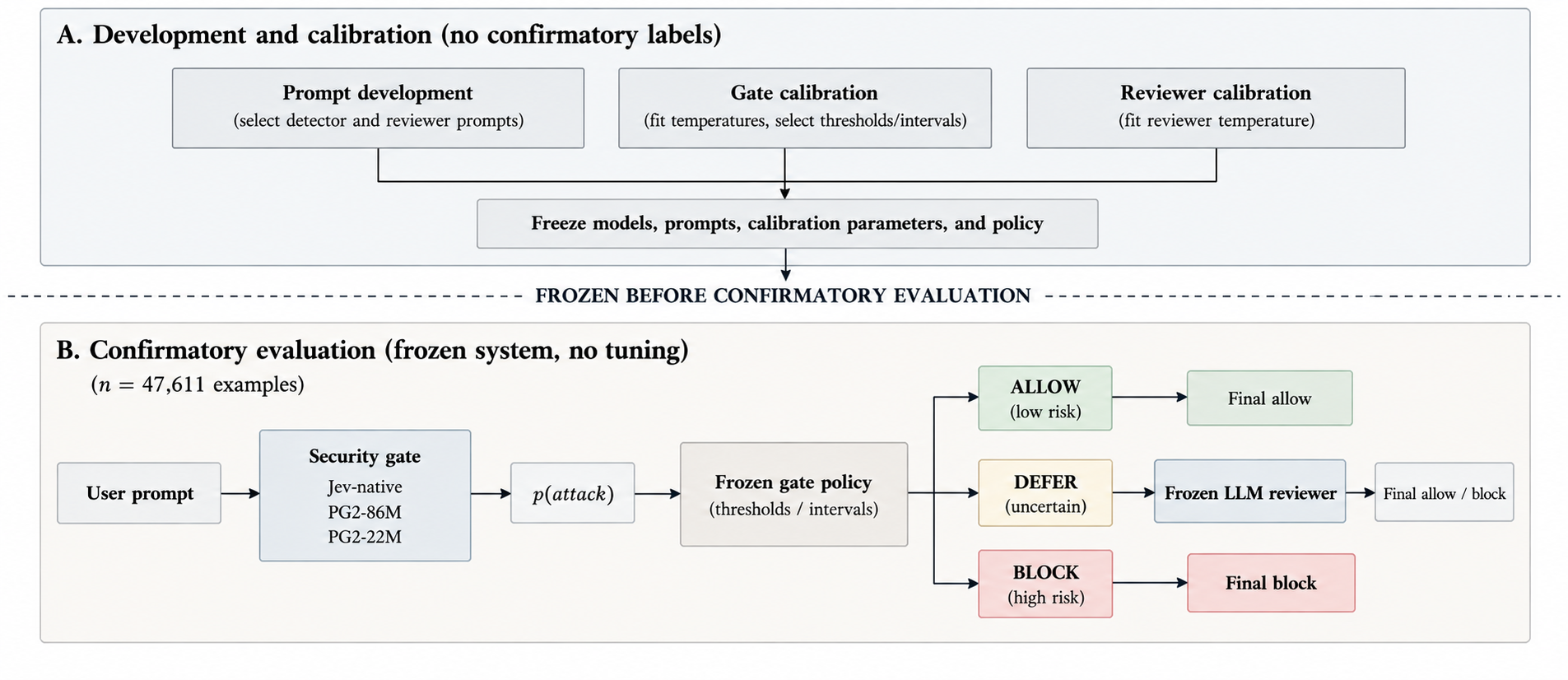}
  \caption{Evaluation design and decision pipeline. (A) Prompts, models, calibration parameters, and the gate policy are selected using development and calibration data only and then frozen. (B) Each confirmatory input is scored by a frozen security gate, and a fixed policy maps the resulting attack probability to ALLOW, DEFER, or BLOCK. Only deferred inputs are sent to the frozen LLM reviewer, which produces the final decision. The confirmatory cohort is evaluated without further tuning.}
  \label{fig:design}
\end{figure}

\section{Related Work}

\paragraph{Prompt-injection evaluation.}
TensorTrust collected more than 126,000 attacks and 46,000 defenses through an online game, showing that human-authored strategies can transfer beyond the game setting~\cite{toyer2024tensortrust}. BIPIA introduced a benchmark for indirect prompt injection through external content and showed that models struggle to distinguish data from actionable instructions~\cite{yi2023bipia}. InjecGuard introduced NotInject to measure over-defense on benign prompts containing injection-related trigger words~\cite{li2024injecguard}. PIDS-Bench broadens this evaluation to hard-benign inputs, obfuscation, and domain and structural shifts under a frozen evaluation protocol~\cite{shire2026pids}.

\paragraph{Calibration and selective prediction.}
Calibration measures how well predicted probabilities correspond to observed frequencies. Temperature scaling adjusts logits using a held-out calibration set while preserving score rankings~\cite{guo2017calibration}. Selective prediction and abstention use confidence to trade automatic coverage for risk~\cite{geifman2017selective}. This standard selective-risk formulation is insufficient for security settings, where false negatives and false positives have different costs and deferred cases are handled by an imperfect reviewer rather than an oracle.

\paragraph{Decision models.}
TypeSafe presents Jev as a System One model that produces typed probabilistic decisions and describes RLCD as a method for training calibrated outputs~\cite{typesafe2026jev}. We test this claim empirically on a frozen external workload, using the native probability returned by the official API without post-hoc calibration. The comparison addresses a deployment-oriented question: whether an off-the-shelf native probability transfers better than a conventional detector calibrated on target data. It does not isolate the effects of the underlying training objectives.

\section{Experimental Design}

\subsection{Data and isolation}

We use four benchmark families selected before the confirmatory run. Table~\ref{tab:data} shows the frozen confirmatory cohort. The pooled test set is deliberately heterogeneous and highly imbalanced, with 41,981 attacks and 5,630 benign examples. We therefore report both pooled and dataset-level metrics and do not report class-conditional metrics when they are undefined for single-class subsets.

\begin{table}[t]
\centering
\caption{Frozen confirmatory cohort.}
\label{tab:data}
\small
\begin{tabular}{lrrr}
\toprule
Dataset & Benign & Attack & Total \\
\midrule
TensorTrust & 0 & 526 & 526 \\
BIPIA & 53 & 36,075 & 36,128 \\
NotInject & 89 & 0 & 89 \\
PIDS-Bench v3 & 5,488 & 5,380 & 10,868 \\
\midrule
Total & 5,630 & 41,981 & 47,611 \\
\bottomrule
\end{tabular}
\end{table}

We separate all development and calibration steps from the held-out confirmatory run. Detector and reviewer prompts are chosen during prompt development. Gate temperatures and ALLOW--DEFER--BLOCK intervals are fitted on a 2,000-example gate-calibration split, while the reviewer temperature is fitted on a separate 1,000-example reviewer-calibration split. We also use a 1,000-example feasibility cohort to check the protocol before the confirmatory run. The calibration and confirmatory sets are disjoint. All models, prompts, calibration parameters, the reviewer threshold, and deployable gate intervals are frozen before confirmatory evaluation. Confirmatory labels are never used to select a deployable threshold. When we report the point at which a full confirmatory deferral curve first reaches a target FNR, we treat it only as a descriptive property of the ranking fixed before evaluation, not as a threshold selected for deployment.

\subsection{Models}

\paragraph{Jev-native.}
We query \texttt{jev-latest} through the official TypeSafe SDK using a fixed prompt-injection question and retain the native probability returned by the model.

\paragraph{Prompt Guard 2.}
We evaluate Llama Prompt Guard 2 22M and 86M~\cite{meta2025promptguard}. For each classifier, we report both the raw attack probability and the probability after scalar temperature scaling fitted on the gate-calibration split. Because temperature scaling is monotonic, the raw and calibrated variants have identical AUROC and AUPRC. They also produce identical decisions at the 0.5 boundary, while their probability-based calibration metrics can differ.

\paragraph{LLM judge.}
We use GPT-5.5 as the LLM judge. We compare three prompts on an independent 1,000-example development split and select one that asks for an attack probability, a binary decision, and a short rationale. The verbalized probabilities are temperature-scaled on a separate 1,000-example reviewer-calibration split, yielding a frozen temperature of 1.591. The FNR-first reviewer operating point is 0.0527. On the calibration split, this threshold yields an FNR of 0 and an FPR of 98.6\%. This trade-off is important for interpreting RQ3.

\subsection{Metrics}

For detection, we report AUROC, AUPRC, TPR at 1\% FPR, and FNR and FPR at the default 0.5 boundary. The results at 0.5 characterize how well the default decision boundary transfers to the target workload; we do not treat this boundary as an optimized operating point. For calibration, we report expected calibration error (ECE), Brier score, negative log-likelihood (NLL), and the mean attack probability assigned to false negatives. ECE is computed using 10 equal-width bins over $[0,1]$, weighted by bin occupancy. All bins are left-closed and right-open except the final bin, which includes 1. The Brier score is the unweighted mean squared error of the predicted probabilities. NLL is the unweighted binary log loss computed with scikit-learn, which internally clips float64 probabilities to machine precision. A low attack probability on a missed attack indicates a confident benign misclassification. We additionally report mean and p95 latency.

Scalar temperature scaling uses $p_{cal}=\sigma(z/T)$, with $T$ fitted by minimizing unweighted binary NLL through bounded scalar optimization over $T\in[0.05,20]$. When native logits are unavailable, probabilities are clipped to $[10^{-7},1-10^{-7}]$ before conversion to logits. The fitting objective applies the same clipping to the transformed probabilities. We use no class or sample weighting. The implementation uses Python 3.9.21, NumPy 2.0.2, SciPy 1.13.1, and scikit-learn 1.6.1.

For RQ3, the gate handles an input automatically unless it falls within the escalation region. Deferred inputs are passed to the frozen LLM reviewer for the final decision, and tied probabilities are moved together. We summarize the frozen ranking using the empirical confirmatory escalation requirement, defined as the smallest escalation rate on the confirmatory deferral curve at which the observed system satisfies $\fnr\leq1\%$. This quantity is a descriptive test-set statistic rather than a deployable threshold selected using confirmatory labels. We also report FPR, automatic coverage, illustrative expected cost under $(C_{FP},C_{FN},C_{review})=(1,100,0.5)$, and latency. Deployable interval policies are selected exclusively on the gate-calibration split. Oracle and fail-closed review are included only as ablations.

\section{Results}

\subsection{RQ1: Detection versus decision}

Table~\ref{tab:detection} shows that the LLM judge achieves the highest pooled AUROC and AUPRC on the confirmatory mixture, while Jev outperforms both Prompt Guard variants on these discrimination metrics. At its native 0.5 decision boundary, however, Jev misses 64.9\% of attacks. The default Prompt Guard boundaries transfer particularly poorly, with both variants missing more than 90\% of attacks. TPR at 1\% FPR is low across all methods, ranging from 0.9\% to 12.5\%. Under the pooled shift mixture, none of the evaluated methods therefore supports autonomous blocking at a low false-positive rate.

\begin{table*}[t]
\centering
\caption{Overall detection performance on the confirmatory cohort.}
\label{tab:detection}
\small
\begin{tabular}{lrrrrr}
\toprule
Model & AUROC & AUPRC & TPR@1\% FPR & FNR@0.5 & FPR@0.5 \\
\midrule
Jev-native & 0.871 & 0.973 & 0.125 & 0.649 & 0.048 \\
PG2-86M & 0.757 & 0.920 & 0.012 & 0.932 & 0.100 \\
PG2-22M & 0.714 & 0.908 & 0.009 & 0.908 & 0.143 \\
LLM judge & 0.897 & 0.977 & 0.084 & 0.171 & 0.130 \\
\bottomrule
\end{tabular}
\end{table*}

The high pooled AUPRC values partly reflect the 88.2\% attack prevalence. BIPIA accounts for most attacks in the pooled cohort, while NotInject contains only benign examples and TensorTrust contains only attacks in this frozen evaluation. The pooled calibration metrics are similarly specific to this deliberately constructed benchmark mixture rather than a natural deployment prevalence. We therefore use the pooled results as a summary of this particular mixture rather than as a universal ranking of the models.

\subsection{RQ2: Trustworthiness of confidence}

Table~\ref{tab:calibration} shows that Jev's native probabilities are not consistently better calibrated than post-hoc calibrated PG2 probabilities on the pooled cohort. Jev has a slightly lower ECE than calibrated PG2-86M (0.541 versus 0.571) and a nearly identical Brier score (0.440 versus 0.439), but a higher NLL (1.432 versus 1.104). The calibrated LLM judge has the lowest pooled NLL, whereas the raw judge has the lowest pooled ECE and Brier score. Temperature scaling substantially improves the Prompt Guard calibration metrics, particularly NLL, without changing discrimination.

\begin{table*}[t]
\centering
\caption{Overall calibration and false-negative confidence. Lower values are better. FN confidence is the mean \pattack{} among false negatives.}
\label{tab:calibration}
\scriptsize
\begin{tabular}{lrrrrr}
\toprule
Probability source & ECE & Brier & NLL & FN confidence & False negatives \\
\midrule
Jev-native & 0.541 & 0.440 & 1.432 & 0.1630 & 27,231 \\
PG2-86M raw & 0.821 & 0.817 & 5.064 & 0.0107 & 39,119 \\
PG2-86M calibrated & 0.571 & 0.439 & 1.104 & 0.2869 & 39,119 \\
PG2-22M raw & 0.801 & 0.794 & 4.625 & 0.0171 & 38,135 \\
PG2-22M calibrated & 0.607 & 0.495 & 1.259 & 0.2339 & 38,135 \\
LLM judge raw & 0.221 & 0.163 & 0.664 & 0.0591 & 7,163 \\
LLM judge calibrated & 0.269 & 0.165 & 0.550 & 0.1306 & 7,163 \\
\bottomrule
\end{tabular}
\end{table*}

Among attacks missed by raw PG2-86M, the mean assigned attack probability is only 0.0107, corresponding to a mean benign probability of 98.9\%. Temperature scaling makes these probabilities less extreme but does not recover the missed attacks at the fixed 0.5 boundary. Jev also assigns low attack probabilities to many missed attacks, with a mean $\pattack=0.163$ among false negatives.

\begin{figure*}[t]
  \centering
  \includegraphics[width=0.9\textwidth]{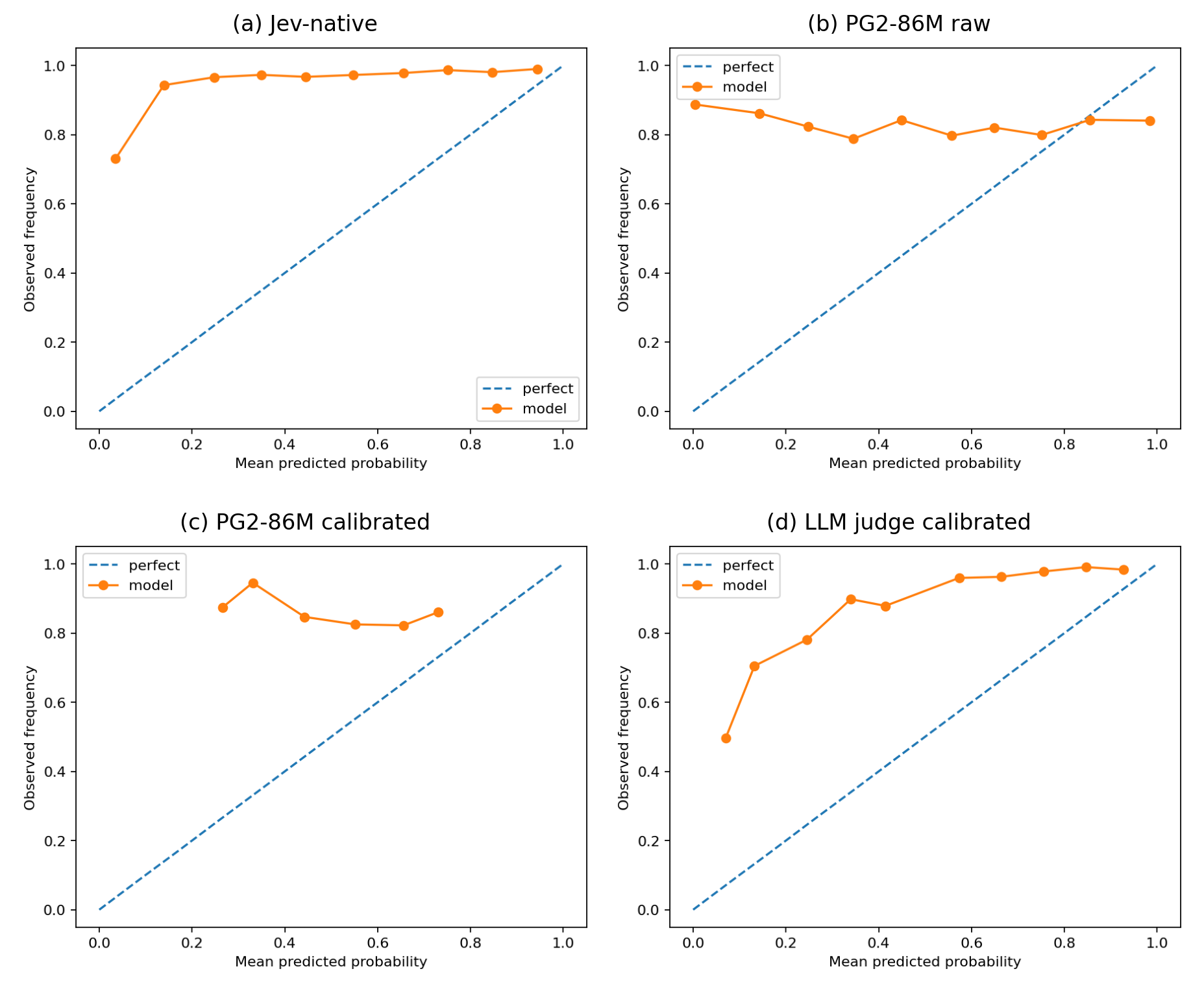}
  \caption{Overall reliability diagrams for the frozen confirmatory mixture, which contains 88.2\% attacks. The diagonal represents perfect calibration. All systems deviate from it under the pooled shift mixture. Raw classifier scores show the largest deviations, while post-hoc scaling reduces but does not eliminate the gap. The curves describe calibration on this benchmark mixture and should not be interpreted as calibration under a natural deployment prevalence.}
  \label{fig:reliability}
\end{figure*}

\subsection{RQ3: ALLOW--BLOCK--DEFER policies}

\begin{figure}[t]
  \centering
  \includegraphics[width=0.9\textwidth]{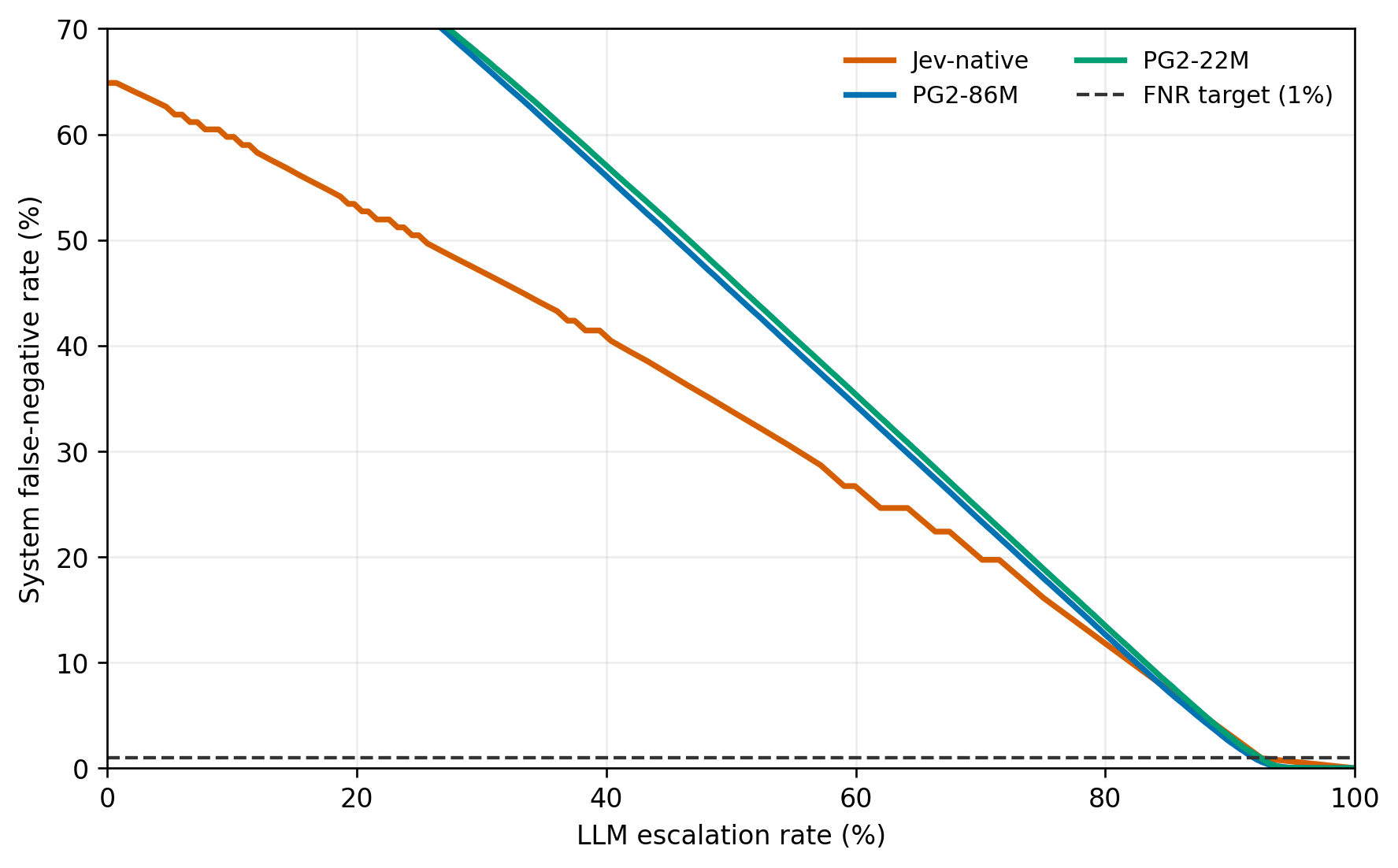}
  \caption{Overall security-deferral curves on the confirmatory cohort, with deferred inputs sent to the frozen LLM reviewer. The dashed line marks the 1\% FNR target. Curve crossings are descriptive properties of rankings fixed before the confirmatory run, not thresholds selected for deployment. Raw and temperature-scaled Prompt Guard curves overlap because scalar temperature scaling preserves uncertainty ordering.}
  \label{fig:deferral}
\end{figure}

Figure~\ref{fig:deferral} shows how system FNR changes as progressively more uncertain cases are escalated to the frozen LLM reviewer. Table~\ref{tab:rq3} reports the empirical confirmatory crossing defined above. Each gate reaches $\fnr\leq1\%$ only after more than 91.8\% of inputs are escalated: 91.89\% for PG2-86M, 92.45\% for PG2-22M, and 92.55\% for Jev. These estimates differ by less than one percentage point. Without uncertainty intervals, we do not interpret these small differences as evidence that one gate is operationally better than another.

\begin{table*}[t]
\centering
\caption{Empirical confirmatory escalation requirement, defined as the first point on each frozen-ranking deferral curve at which the observed overall system satisfies $\fnr\leq1\%$. This is a descriptive test statistic rather than a deployment threshold selected using confirmatory labels. Raw and calibrated PG2 rows coincide because temperature scaling preserves uncertainty ordering.}
\label{tab:rq3}
\scriptsize
\begin{tabular}{lrrrrrr}
\toprule
Gate & Escalation & Auto. coverage & System FNR & System FPR & Norm. cost & Mean latency (ms) \\
\midrule
PG2-86M raw/cal. & 91.89\% & 8.11\% & 0.998\% & 40.55\% & 1.387 & 3,713 \\
PG2-22M raw/cal. & 92.45\% & 7.55\% & 0.998\% & 43.69\% & 1.394 & 3,734 \\
Jev-native & 92.55\% & 7.45\% & 0.943\% & 45.84\% & 1.349 & 3,885 \\
\bottomrule
\end{tabular}
\end{table*}

The system FPR remains between 40.6\% and 45.8\% at these crossings, showing why FNR alone is not sufficient to characterize the resulting policy. The reviewer threshold was selected under an FNR-first calibration objective and has an FPR of 98.6\% on the reviewer-calibration split. Escalating most inputs to this reviewer reduces false negatives but also causes many benign inputs to be blocked. Under the frozen protocol, this result shows that probability calibration alone does not determine a useful decision policy. Reviewer behavior and the security and utility objectives remain important properties of the deployed system.

\subsection{Latency and deployment trade-offs}

\begin{table}[t]
\centering
\caption{Observed per-example latency in the evaluated setup.}
\label{tab:latency}
\small
\begin{tabular}{lrrr}
\toprule
Model & Mean (ms) & p50 (ms) & p95 (ms) \\
\midrule
PG2-86M & 43.8 & 38.4 & 69.7 \\
PG2-22M & 45.8 & 45.3 & 76.1 \\
Jev-native & 190.9 & 155.2 & 301.7 \\
LLM judge & 3,890.2 & 3,476.0 & 6,603.9 \\
\bottomrule
\end{tabular}
\end{table}

The local Prompt Guard models are roughly two orders of magnitude faster than the remote LLM judge. Jev is slower than the local classifiers but substantially faster than the reviewer. Because the API measurements include network and provider overhead, these latency values characterize the evaluated deployment setup rather than model-only computation.

\section{Discussion}

\paragraph{A calibrated decision objective does not guarantee reliable deployment.}
Jev's native probabilities are better calibrated than raw Prompt Guard scores on several pooled metrics, but they are not consistently better than those of a conventional classifier after temperature scaling on target calibration data. Calibration therefore needs to be evaluated on the target workload rather than inferred from the model's training objective.

\paragraph{Post-hoc calibration adjusts confidence, not representation.}
Temperature scaling substantially improves PG2's ECE, Brier score, and NLL, but leaves AUROC, AUPRC, decisions at the 0.5 boundary, and uncertainty ordering unchanged. Under substantial distribution shift, calibration cannot compensate for information that the classifier has failed to capture.

\paragraph{Reviewer quality constrains the benefit of deferral.}
A weak reviewer, or one operating at an aggressive threshold, limits what deferral can achieve. Our reviewer has the strongest discrimination among the evaluated models, but its FNR-first operating point also produces a very high FPR. As the escalation rate approaches 100\%, overall system behavior converges toward reviewer behavior. The deferral curves therefore characterize the gate--reviewer system as a whole rather than the gate alone.

\paragraph{Pooled metrics can obscure composition effects.}
The confirmatory cohort is dominated by attacks from BIPIA. ECE and Brier score depend on the evaluation distribution, while AUPRC is sensitive to class prevalence. The pooled results therefore characterize this frozen benchmark mixture rather than a natural deployment distribution. We accordingly limit our calibration conclusions to this mixture and report dataset-level calibration separately.

\section{Threats to Validity and Limitations}

Jev is a hosted model evaluated using a single fixed question and API version, so different wording or later model versions may yield different results. The LLM judge produces verbalized probabilities, which are not equivalent to probabilities derived from token-level likelihoods. The confirmatory mixture does not reflect a natural production prevalence, and its pooled calibration and cost results should not be transferred directly to deployment settings.

Some dataset subsets contain only one class, making certain class-conditional metrics undefined. For these subsets, we report only metrics that remain meaningful, such as false-positive behavior. The cost model $(1,100,0.5)$ is illustrative, and deployment policies should instead be selected using application-specific costs and review budgets. Latency is measured in the evaluated combination of local and API-based inference without normalization across controlled hardware.

The comparison also differs in how the models are adapted to the target workload. PG2 is temperature-scaled on target calibration data, whereas the primary Jev evaluation intentionally uses its native probability without post-hoc calibration. A post-hoc calibration ablation for Jev would help distinguish the transfer of its native calibration from its ability to be calibrated after adaptation.

We report point estimates without confidence intervals. Benchmark transformations may also introduce dependence among examples, so small numerical differences, particularly the sub-percentage-point differences in RQ3, should not be treated as statistically resolved. Finally, our evaluation focuses on detection and routing rather than end-to-end agent compromise or downstream harm.

\section{Conclusion}

We evaluated whether a model explicitly trained for calibrated decisions produces more reliable prompt-injection security decisions than specialized classifiers and an LLM judge. Across 47,611 held-out examples, Jev ranks attacks better than Prompt Guard 2 but is not consistently better calibrated than temperature-scaled PG2-86M. The LLM judge provides the strongest discrimination and overall probability estimates among the evaluated methods, but at substantially higher latency. On the confirmatory deferral curves, every tested gate reaches an observed $\fnr\leq1\%$ only after more than 91.8\% of inputs are escalated, while still producing substantial false positives. These crossings are descriptive test statistics rather than deployment thresholds selected using confirmatory labels. Our results show that training for calibrated outputs alone is not sufficient for reliable deployment decisions. Discrimination, calibration, reviewer behavior, and policy utility need to be evaluated together as parts of the same decision pipeline.

\bibliographystyle{plain}
\bibliography{references}

\appendix
\section{Dataset-level Calibration}

\begin{table*}[h]
\centering
\caption{Calibration by dataset for Jev and PG2-86M. Single-class datasets retain valid proper-score summaries but not all discrimination metrics.}
\label{tab:dataset-calibration}
\small
\begin{tabular}{llrrr}
\toprule
Model & Dataset & ECE & Brier & NLL \\
\midrule
Jev-native & BIPIA & 0.660 & 0.530 & 1.723 \\
Jev-native & NotInject & 0.045 & 0.009 & 0.052 \\
Jev-native & PIDS-Bench v3 & 0.167 & 0.162 & 0.520 \\
Jev-native & TensorTrust & 0.204 & 0.136 & 0.484 \\
\midrule
PG2-86M raw & BIPIA & 0.984 & 0.979 & 6.157 \\
PG2-86M raw & NotInject & 0.070 & 0.047 & 0.257 \\
PG2-86M raw & PIDS-Bench v3 & 0.319 & 0.319 & 1.673 \\
PG2-86M raw & TensorTrust & 0.179 & 0.165 & 0.886 \\
\midrule
PG2-86M calibrated & BIPIA & 0.713 & 0.512 & 1.262 \\
PG2-86M calibrated & NotInject & 0.295 & 0.098 & 0.366 \\
PG2-86M calibrated & PIDS-Bench v3 & 0.118 & 0.214 & 0.616 \\
PG2-86M calibrated & TensorTrust & 0.342 & 0.143 & 0.463 \\
\bottomrule
\end{tabular}
\end{table*}

\section{Frozen Policy Artifacts}

The reviewer prompt is \texttt{llm-judge-candidate-effect-v1}. Its scalar
temperature is 1.590979 and its frozen calibrated-probability threshold is
0.052738. Gate ALLOW/DEFER/BLOCK intervals are selected only on the
2,000-example gate calibration split. The primary security--deferral curves use
the real frozen reviewer decision rule. Fixed intervals with oracle and
fail-closed review outcomes are ablations and are not interpreted as
deployed-system performance.

\section{Reproducibility Artifacts}
The code and reproducibility artifacts are available at \url{https://github.com/lanqianlong/calibrated-decision-models}. All reported rows are generated from exact-ID joins. The released prediction
files contain 47,611 aligned records for Jev-native, PG2-22M raw/calibrated,
PG2-86M raw/calibrated, and LLM judge raw/calibrated; the RQ3 pipeline also
uses the frozen LLM-judge operating-point predictions. The artifact includes
\texttt{paper\_metrics.json}, a compact \texttt{security\_deferral.json}
reference containing the primary endpoints and operating-point ablations, and
CSV/figure outputs regenerated by \texttt{scripts/reproduce\_all.sh}. Full
security--deferral curves are regenerated offline from the bundled prediction
artifacts rather than stored as a large reference JSON.

\end{document}